# Quenching of star formation from a lack of inflowing gas to galaxies

Katherine E. Whitaker[1,2], Christina C. Williams[3], Lamiya Mowla[4], Justin S. Spilker[5,6], Sune Toft[2,7], Desika Narayanan[8,2], Alexandra Pope[1], Georgios E. Magdis[2,9,7], Pieter G. van Dokkum[10], Mohammad Akhshik[11], Rachel Bezanson[12], Gabriel B. Brammer[2,7], Joel Leja[13,14,15], Allison Man[16], Erica J. Nelson[17], Johan Richard[18], Camilla Pacifici[19], Keren Sharon[20], Francesco Valentino[2,7]

**Star formation in half of massive galaxies was quenched by the time the Universe was three billion years old[1]. Very low amounts of molecular gas appear responsible for this, at least in some cases[2-7], though morphological gas stabilization, shock heating, or activity associated with accretion onto a central supermassive black hole is invoked in other cases[8-11]. Recent studies of quenching by gas depletion have been based upon upper limits that are insufficiently sensitive to determine this robustly[2-7], or stacked emission with its problems of averaging[8,9]. Here we report 1.3mm observations of dust emission from six strongly lensed galaxies where star formation has been quenched, with magnifications of up to a factor of 30. Four of the six galaxies are undetected in dust emission, with an estimated upper limit on the dust mass of 0.0001 times the stellar mass, and by proxy (assuming a Milky Way molecular gas-to-dust ratio) 0.01 times the stellar mass in molecular gas. This is two orders of magnitude less molecular gas per unit stellar mass than seen in star forming galaxies at similar redshifts[12-14]. It remains difficult to extrapolate from these small samples, but these observations establish that gas depletion is responsible for a cessation of star formation in some fraction of high-redshift galaxies.**

[1] Department of Astronomy, University of Massachusetts, Amherst, MA, 01003 USA
[2] Cosmic Dawn Center (DAWN), Copenhagen, Denmark
[3] Steward Observatory, University of Arizona, 933 North Cherry Avenue, Tucson, AZ 85721, USA
[4] Dunlap Institute for Astronomy and Astrophysics, University of Toronto, 50 St George St, Toronto, ON M5S 3H4, Canada
[5] Department of Astronomy, University of Texas at Austin, 2515 Speedway, Stop C1400, Austin, TX 78712, USA
[6] NASA Hubble Fellow
[7] Niels Bohr Institute, University of Copenhagen, Jagtvej 128, DK2200, Copenhagen, Denmark
[8] Department of Astronomy, University of Florida, 211 Bryant Space Sciences Center, Gainesville, FL 32611 USA
[9] DTU-Space, Technical University of Denmark, Elektrovej 327, DK-2800 Kgs. Lyngby, Denmark
[10] Astronomy Department, Yale University, 52 Hillhouse Ave, New Haven, CT 06511, USA
[11] Department of Physics, University of Connecticut, Storrs, CT 06269, USA
[12] Department of Physics and Astronomy and PITT PACC, University of Pittsburgh, Pittsburgh, PA 15260, USA
[13] Department of Astronomy & Astrophysics, The Pennsylvania State University, University Park, PA 16802, USA
[14] Institute for Computational & Data Sciences, The Pennsylvania State University, University Park, PA, USA
[15] Institute for Gravitation and the Cosmos, The Pennsylvania State University, University Park, PA 16802, USA
[16] Department of Physics & Astronomy, University of British Columbia, 6224 Agricultural Road, Vancouver, BC V6T 1Z1, Canada
[17] Department of Astrophysical and Planetary Sciences, 391 UCB, University of Colorado, Boulder, CO 80309-0391, USA
[18] Univ Lyon, Univ Lyon1, Ens de Lyon, CNRS, Centre de Recherche Astrophysique de Lyon UMR5574, F-69230, Saint-Genis-Laval, France
[19] Space Telescope Science Institute, Baltimore, MD 21218, USA
[20] Department of Astronomy, University of Michigan, 1085 S. University Ave, Ann Arbor, MI 48109, USA

The 1.3mm observations were made with the *Atacama Large Millimeter/submillimeter Array (ALMA)* and the sample comprises of six galaxies selected from the REsolving QUIEscent Magnified (REQUIEM) galaxy survey: MRG-M1341[15], MRG-M0138[16], MRG-M2129[17], MRG-M0150[16], MRG-M0454[18], and MRG-M1423[18] (Figure 1). The targets are all strongly lensed, with magnification factors ranging from a factor of 2.7 (MRG-M1423) to 30 (MRG-M1341). Five out of the six galaxies are classified as quiescent due to unusually low star-formation rates that reach down to 0.1 $M_\odot yr^{-1}$, as measured from fitting the optical to infrared spectral energy distributions (see Methods). While the most distant target, MRG-M1423, has a more typical star-formation rate of ~140 $M_\odot yr^{-1}$ over the previous 100 Myr, consistent with normal star-forming galaxies at $z = 3$, its spectrum reveals classic post-starburst signatures that support a picture where it has quenched rapidly within the last 100 Myr [18]. These targets are qualitatively different than existing millimeter/carbon monoxide (CO) data tracing cold interstellar medium phases in quenched galaxies in that they have order of magnitude lower star-formation rates for their stellar mass[2,4-6,8,11], higher redshifts[3,7,9,10], and uniquely deep flux limits facilitated by strong lensing magnification.

For the redshift range of our sample, our 1.3mm wavelength observations correspond to 300-500μm rest-frame on the Rayleigh-Jeans tail of the dust emission, which serves as a robust proxy for the cold molecular gas mass[19]. We clearly detect two of the sources in the dust continuum: MRG-M0138 at 0.27±0.03 mJy and MRG-M2129 at 9.74±0.16 mJy. Such direct detections of cold dust in individual quiescent galaxies outside the local universe, implying percent-level molecular gas fractions, are scant due to the extreme sensitivity requirements. In contrast with the extended stellar light profiles, and despite the enhanced resolution from strong lensing magnification, both sources remain unresolved with no evidence for missing extended flux (see Methods). This suggests that they have high dust and molecular gas surface densities, as the dust continuum is centrally concentrated and significantly less extended than the stellar light (see Figure 1). Such a result has also been found in star-forming galaxies at similar redshifts[20]. The sensitive *ALMA* dust continuum imaging of the remaining four sources all yield strong upper limits, with the 3σ detection limits ranging from 30-150 μJy before lensing corrections. We estimate the dust mass, $M_{dust}$, by adopting a modified blackbody fit and making standard assumptions about dust temperature and emissivity (see Methods).

We show the redshift evolution of the dust fraction, $f_{dust}=M_{dust}/M_\star$, for our sample of lensed quenched galaxies in Figure 2. By adopting a ratio of the molecular gas mass to dust mass of 100 (see Methods), we estimate $M_{H2}$ directly from $M_{dust}$ and also show the inferred molecular gas fraction, $f_{H2} = M_{H2}/M_\star$ (right axis). Even if we adopt an extremely conservative molecular gas to dust mass ratio that is a factor of ten higher, $f_{H2}$ is still well below that of normal star-forming galaxies at this epoch[14]. Both of our unambiguously detected galaxies have low molecular gas fractions of 4.6±0.5% and 0.6±0.1%, respectively, with systematic uncertainties in dust temperature and molecular gas mass to dust mass ratio of a factor of 1.7. Strong upper limits from CO emission for these two targets rule out more exotic molecular gas-to-dust ratios in these particular cases, which would otherwise imply larger cold gas reservoirs (see Methods). While scaling relations adequately describe the cold gas content of quiescent galaxies in the local Universe by construction[21] (e.g., contours in Figure 2), our observations reveal a population of massive galaxies at $z > 1.5$ that have molecular gas fractions more than order of magnitude lower than empirical predictions at similar redshifts. Our measured $f_{H2}$ is 0.9±0.2 dex lower on average than scaling relation predictions for the given star-formation rates and stellar mass[14].

Our program measures a broad range of (low) molecular gas masses in massive galaxies with suppressed star-formation rates (Figure 3). A comprehensive literature search at $1.5<z<3.0$ (see Methods) demonstrates that galaxies typically form copious new stars (median specific star formation rate, log(SFR/$M_\star$) = -8.6) and have a bountiful fuel supply, with a median value of $f_{H2}$ = 51%. By comparison, our galaxies instead form two orders of magnitude fewer new stars (median log(SFR/$M_\star$) = -10.7) and have a median upper limit of $f_{H2}$ < 1%. Until recently, such low molecular gas fractions have only been measurable in galaxies in the local universe[21]. Our new measurements confirm that the cold interstellar medium was already rapidly depleted at high redshift in at least some galaxies, not slowly consumed until the present day.

Another study has already set the stage at high redshift, finding moderate cold gas reservoirs based on stacking dust continuum measurements in a mass-representative sample[8]. While the cold gas reservoir of MRG-M2129 is consistent with these first results, all other sources remain in significant tension. The sample is too small for us to distinguish if we are sampling a biased sub-population, or if contamination due to the significantly lower resolution of the earlier stacking study biases earlier measurements high. Our results also contradict the moderate cold gas reservoirs detected in recently-quenched galaxies at lower redshifts that instead imply reduced star formation efficiency [10]. While in principle, differences in the ages of the stellar populations could explain this discrepancy[22], our sample includes both recently-quenched (~100-800 Myr) and older passively evolving galaxies (~1.3-1.6 Gyr)[15-18]. Future observations constraining the distribution of dust temperatures may add clarity to these differences: because dust temperature changes the peak wavelength of the FIR dust bump, an overall hotter average dust temperature will decrease the mm flux density for a given total IR luminosity, whereas it will increase for colder dust temperatures[9]. The large range in molecular gas and dust fractions observed at low star-formation rate across redshift, from <2-5%[2-7] to 10%[8,9] up to 40%[3,10,11], may suggest a diversity in dust temperatures (see Methods), or, more fundamentally, a diverse set of evolutionary pathways to quiescence. The galaxies in our sample either depleted their cold gas within the first few billion years of the Big Bang, or ejected it into the surrounding intergalactic medium. Chemical evolution arguments based on observed high metallicity and high alpha/Fe ratios in local early-type galaxies support the same interpretation, where massive galaxies must have consumed all of the available gas within roughly one billion years [23]. Larger samples to similar or greater depth are needed to determine if this scenario is generally applicable.

Quiescent galaxies are spectroscopically confirmed as early as $z=4$[24]. The existence of these early quiescent galaxies and the rapid and complete exhaustion of gas implied by our data are critical constraints on models of galaxy evolution, which currently struggle to produce realistic quiescent galaxies across redshift [21]. Predictions from cosmological simulations for the molecular gas leftover after star formation ceases span multiple orders of magnitude [25,26]. The essential problem is that high redshift dark matter halos contain enormous gas reservoirs[12-14] that should cool efficiently and maintain steady star formation over long timescales[27,28]. Indeed, many early massive galaxies do just that, having star-formation rates of order 100 $M_\circ yr^{-1}$ [29] and sizable molecular gas reservoirs[13]. Our new observations show that the cessation of star formation for these galaxies is not caused by a sudden inefficiency in the conversion of cold gas to stars but due to the depletion or removal of their reservoirs.

This lack of cold gas may be permanent. In the absence of a heating mechanism, the hot gas biding time in the halo of massive galaxies should theoretically cool and fall back onto galaxies within a billion years[30]. Yet, we do not frequently observe rejuvenation in massive galaxies[31]. In light of this, there must be a physical mechanism that effectively blocks the replenishment of the cold gas reservoirs[32]. In the local Universe, centrally driven winds observed in quiescent galaxies are known to clear the gas out of the system, and the central low-level active supermassive black hole has sufficient mechanical energy to heat the gas and suppress star formation[33]. Tentative evidence also exists at high redshifts for maintenance mode energy injection from central supermassive black holes[34]. This process may explain why quiescent galaxies are unable to effectively re-accrete cold gas in the subsequent 10 billion years of evolution to the present day, although there are other possibilities[35]. Our new data demonstrate a lack of dust, and by inference cold gas, indicating that such a physical process may have already occurred at significantly earlier times for some galaxies.

With extremely sensitive limits on the dust continuum of individual massive quiescent galaxies at $z\sim2$, our measurements imply extremely low $f_{H2}$ of a few percent or less. However, the use of the dust continuum as a proxy for the interstellar medium in massive galaxies with star-formation rates must be further investigated. In particular, while securing detections of both CO emission and dust continuum for the same high redshift quiescent galaxy is paramount, such observations are costly with our current generation of telescopes without the help of strong gravitational lensing magnification.

**References**

1. Muzzin, A., et al., "The Evolution of the Stellar Mass Functions of Star-forming and Quiescent Galaxies to z=4 from the COSMOS/UltraVISTA Survey", Astrophys. J., 777, 18 (2013).
2. Sargent, M. et al., "A Direct Constraint on the Gas Content of a Massive, Passively Evolving Elliptical Galaxy at $z = 1.43$", Astrophys. J., 806, 20 (2015).
3. Spilker, J., et al., "Molecular Gas Contents and Scaling Relations for Massive, Passive Galaxies at Intermediate Redshifts from the LEGA-C Survey", Astrophys. J., 860, 103 (2018).

4. Bezanson, R., et al., "Extremely Low Molecular Gas Content in a Compact, Quiescent Galaxy at z = 1.522", Astrophys. J., 873, 19 (2019).
5. Zavala, J., et al., "On the Gas Content, Star Formation Efficiency, and Environmental Quenching of Massive Galaxies in Protoclusters at z ~ 2.0-2.5", Astrophys. J., 887, 183 (2019).
6. Caliendo, J., et al., "Early Science with the Large Millimeter Telescope: Constraining the Gas Fraction of a Compact Quiescent Galaxy at z=1.883", Astrophys. J. L., 910, 7 (2021).
7. Williams, C., et al., "ALMA measures rapidly depleted molecular gas reservoirs in massive quiescent galaxies at *z*~1.5", Astrophys. J., 908, 54 (2021).
8. Gobat, R., et al., "The unexpectedly large dust and gas content of quiescent galaxies at z>1.4", Nature Astron., 2, 239-246 (2018).
9. Magdis, G., et al., " The Interstellar Medium of Quiescent Galaxies and its Evolution With Time", Astron. Astrophys., 647, 33 (2021).
10. Suess, K., et al., "Massive Quenched Galaxies at z~0.7 Retain Large Molecular Gas Reservoirs", Astrophys. J., 846, 14 (2017).
11. Hayashi, M., et al., "Molecular Gas Reservoirs in Cluster Galaxies at z = 1.46", Astrophys. J., 856, 118 (2018).
12. Tacconi, L., et al. "High molecular gas fractions in normal massive star-forming galaxies in the young Universe", Nature, 463, 781-784 (2010).
13. Genzel, R., et al., "Combined CO and Dust Scaling Relations of Depletion Time and Molecular Gas Fractions with Cosmic Time, Specific Star-formation Rate, and Stellar Mass", Astrophys. J., 800, 20 (2015).
14. Tacconi, L., et al., "PHIBSS: Unified Scaling Relations of Gas Depletion Time and Molecular Gas Fractions", Astrophys. J., 853, 179 (2018).
15. Ebeling, H., et al., "Thirty-fold: Extreme Gravitational Lensing of a Quiescent Galaxy at z=1.6", Astrophys. J., 852, 7 (2018).
16. Newman, N., et al., "Resolving Quiescent Galaxies at z>2. I. Search for Gravitationally Lensed Sources and Characterization of Their Structure, Stellar Populations, and Line Emission", Astrophys. J., 862, 125 (2018).
17. Toft, S., et al., "A massive, dead disk galaxy in the early Universe", Nature, 546, 510-513 (2017).
18. Man, A., et al., "An exquisitely deep view of quenching galaxies through the gravitational lens: Stellar population, morphology, and ionized gas", (in press) Astrophys. J., arXiv:2106.08338.
19. Scoville, N., et al., "ISM Masses and the Star formation Law at Z = 1 to 6: ALMA Observations of Dust Continuum in 145 Galaxies in the COSMOS Survey Field", Astrophys. J., 820, 83 (2016).
20. Tadaki, K., et al., "Bulge-forming Galaxies with an Extended Rotating Disk at z ~ 2", Astrophys. J., 824, 175 (2017).
21. Saintonge, A., et al., "xCOLD GASS: The Complete IRAM 30 m Legacy Survey of Molecular Gas for Galaxy Evolution Studies", Astrophys. J. Supplement, 233, 22 (2017).
22. Li, Z., et al., "The Evolution of the Interstellar Medium in Post-starburst Galaxies", Astrophys. J.. 879, 131 (2019).
23. Thomas, D, et al., "The Epochs of Early-Type Galaxy Formation as a Function of Environment", Astrophys. J., 621, 673 (2005).
24. Valentino, F., et al., "Quiescent Galaxies 1.5 Billion Years after the Big Bang and Their Progenitors", Astrophys. J., 889, 93 (2020).

25. Lagos, C., et al., "The origin of the atomic and molecular gas contents of early-type galaxies - II. Misaligned gas accretion", Mon. Not. R. Astron. Soc., 448, 1271-1287 (2015).
26. Dave, R., et al., "SIMBA: Cosmological simulations with black hole growth and feedback", Mon. Not. R. Astron. Soc., 486, 2827-2849 (2019).
27. Keres, D., et al., "How do galaxies get their gas?, Mon. Not. R. Astron. Soc., 363, 2-28 (2005).
28. Dekel, A., et al. "Cold streams in early massive hot haloes as the main mode of galaxy formation", Nature, 457, 451-454 (2009).
29. Whitaker, K., et al., "Constraining the Low-mass Slope of the Star Formation Sequence at $0.5 < z < 2.5$", Astrophys. J., 775, 104 (2014).
30. Ciotti, L., et al., "Radiative Feedback from Massive Black Holes in Elliptical Galaxies: AGN Flaring and Central Starburst Fueled by Recycled Gas", Astrophys. J., 665, 1038-1056 (2007).
31. Akhshik, M., et al., "Recent Star-formation in a Massive Slowly-Quenched Lensed Quiescent Galaxy at z=1.88", Astrophys. J. L., 907, 8 (2021).
32. Dekel, A. & Birnboim, Y., "Galaxy bimodality due to cold flows and shock heating", Mon. Not. R. Astron. Soc., 368, 2-20 (2006).
33. Cheung, E., et al., "Suppressing star formation in quiescent galaxies with supermassive black hole winds", Nature, 533, 504-508 (2016).
34. Whitaker, K., et al., "Quiescent Galaxies in the 3D-HST Survey: Spectroscopic Confirmation of a Large Number of Galaxies with Relatively Old Stellar Populations at z~2", Astrophys. J. L., 770, 39 (2013).
35. Johansson, P., et al., "Gravitational Heating Helps Make Massive Galaxies Red and Dead", Astrophys. J. L. , 697, L38-L43 (2009).

**Acknowledgements:** This paper makes use of the following ALMA data: ADS/JAO.ALMA #2018.1.00276.S, ADS/JAO.ALMA #2019.1.00227.S. ALMA is a partnership of ESO (representing its member states), NSF (USA) and NINS (Japan), together with NRC (Canada), MOST and ASIAA (Taiwan), and KASI (Republic of Korea), in cooperation with the Republic of Chile. The Joint ALMA Observatory is operated by ESO, AUI/NRAO and NAOJ. The National Radio Astronomy Observatory is a facility of the National Science Foundation operated under cooperative agreement by Associated Universities, Inc. This work uses observations with the NASA/ESA Hubble Space Telescope, obtained at the Space Telescope Science Institute, which is operated by the Association of Universities for Research in Astronomy, Inc., under NASA contract NAS 5-26555. KEW wishes to acknowledge funding from the Alfred P. Sloan Foundation, HST-GO-14622, HST-GO-15663, and the unwavering support of her partner amid the global pandemic. CCW acknowledges support from the National Science Foundation Astronomy and Astrophysics Fellowship grant AST-1701546 and from NIRCam Development Contract NAS50210 from NASA Goddard Space Flight Center to the University of Arizona. S.T. acknowledges support from the ERC Consolidator Grant funding scheme (project ConTExt, grant No.648179), FV from the Carlsberg Foundation Research Grant CF18-0388, and GEM from the Villum Fonden research grant 13160. The Cosmic Dawn Center is funded by the Danish National Research Foundation under grant No. 140. CP is supported by the Canadian Space Agency under a contract with NRC Herzberg Astronomy and Astrophysics. MA acknowledges support by NASA under award No 80NSSC19K1418. JSS is supported by NASA Hubble Fellowship grant #HF2-51446 awarded by the Space Telescope Science Institute, which is operated by the Association of Universities for Research in Astronomy, Inc., for NASA, under contract NAS5-26555. AM is supported by a Dunlap Fellowship at the Dunlap Institute for Astronomy & Astrophysics, funded through an endowment established by the David Dunlap family and the University of Toronto.

**Author Contributions:** KEW proposed and carried out the observations, conducted the analysis in this paper, and authored the majority of the text. CCW performed the weighted stack of the data, helped to create Figure 2 and 3, and edited the text in the main body. LM performed the direct analysis of the ALMA flux densities and created the images in Figure 1. JSS carried out the reduction and direct analysis of the raw ALMA data. MA reduced the HST images and MA and JL performed a stellar population synthesis analysis. GEM, AP, ST, and FV helped interpret the millimeter data and contributed to the dust and gas mass analysis. DN helped interpret the data in the context of cosmological simulation models. All authors, including RB, GBB, JL, AM, EJN, CP, KS, and PvD, contributed to the overall interpretation of the results and various aspects of the analysis and writing.

**Author Information:** Please send correspondence to Katherine E. Whitaker, kwhitaker@astro.umass.edu.

The authors have no competing financial interests.

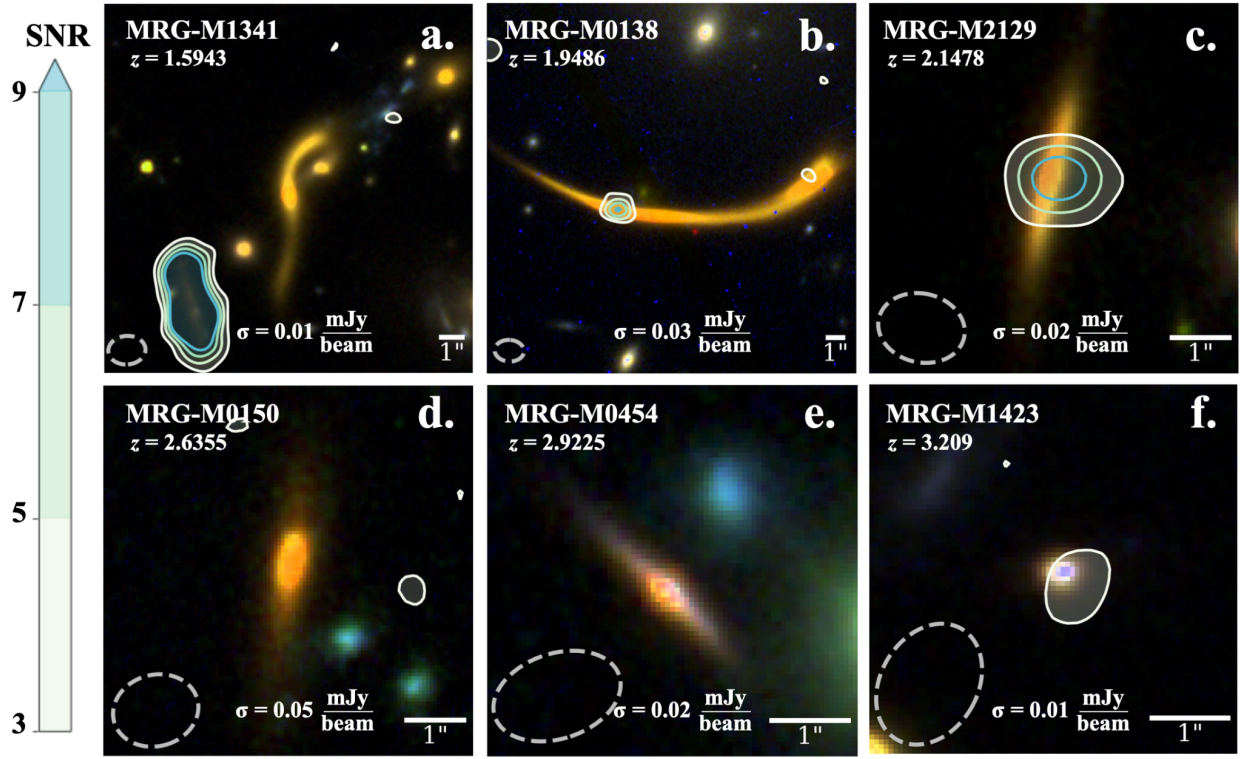

**Figure 1 | Images of six massive lensed galaxies where star formation has been quenched.** The panels are rank-ordered from $z=1.6$ to $z=3.2$ (labeled a-f), showing a composite *Hubble Space Telescope* (*HST*) color image ($i_{F814W}$, $J_{F125W}$, $H_{F160W}$ generally, substituting $J_{F110W}$ for panel e) and contours of ALMA/Band 6 dust continuum observations. Each image is centered on the target galaxy, whose redshift is listed in the upper-left corner. The dashed ellipse indicates the ALMA beam size, with the 1σ noise level noted at the bottom of each panel in units of mJy per beam.

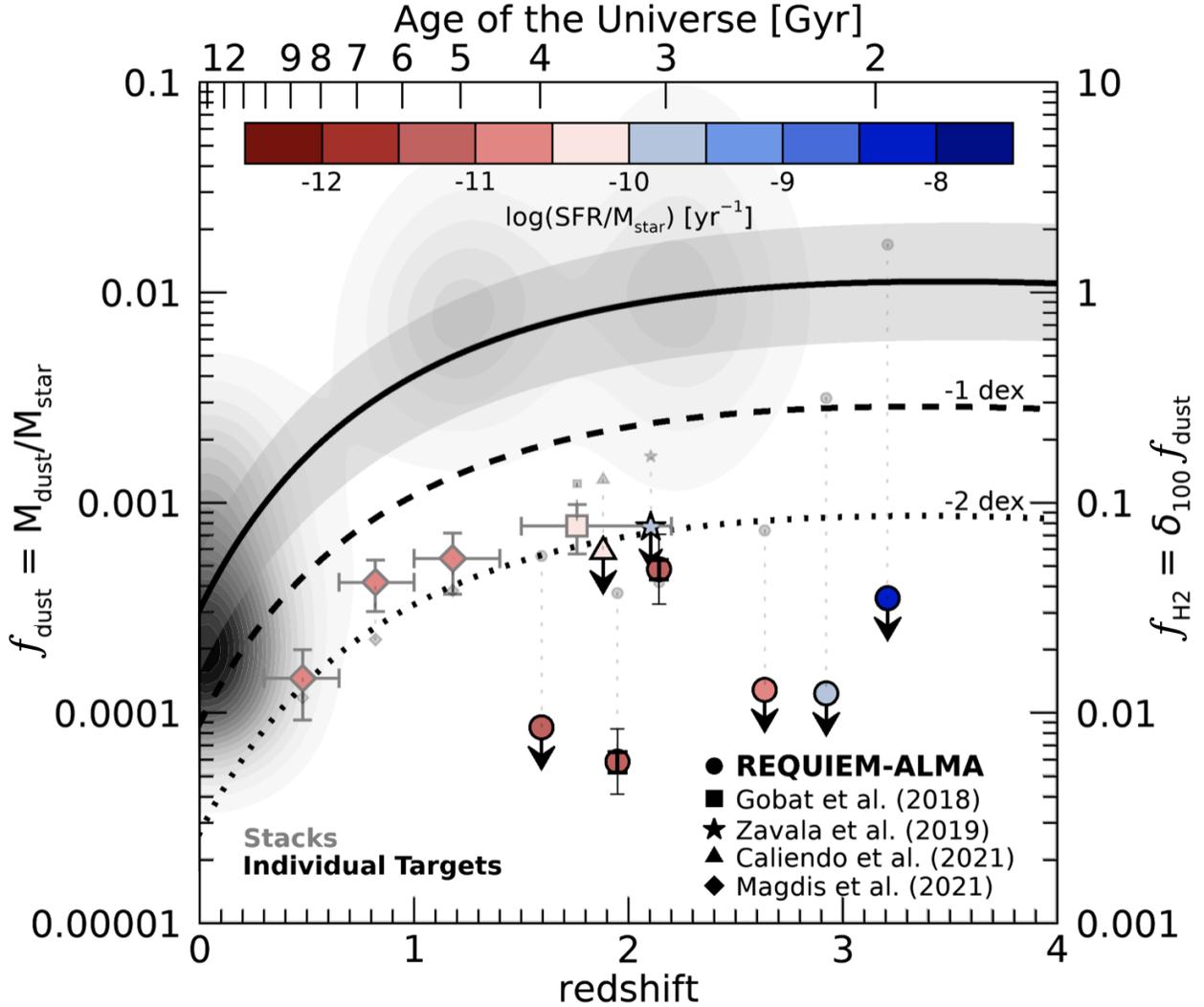

**Figure 2 | Low dust masses for quenched galaxies.** Measurements of $f_{dust}$ for distant lensed quiescent galaxies (circles) are extremely low given their star-formation rate per unit stellar mass (sSFR). We compare existing dust continuum measurements in the literature of individual quiescent galaxies at $z>1.5$[5,6] (individual black symbols) and stacked quiescent galaxies[8,9] from JCMT/SCUBA and ASTE/AzTEC data out to $z\sim2$ (large grey symbols), using identical conversions herein to our sample (see Methods). The thick black error bars are the formal $1\sigma$ measurement uncertainty in our 1.3mm flux density and the thin black error bars represent systematic uncertainties when varying dust temperature. The smaller transparent symbols represent the predicted $f_{H2}$ from empirical scaling relations[4] given sSFR. The inferred $f_{H2}$ (right axis) and scaling relations[14] for $\log(M_\star/M_\odot)=11$ on the average $\log(SFR)$-$\log(M_\star)$ relation (solid), 1 dex (dashed), and 2 dex below (dotted) assume a molecular gas to dust mass ratio of 100. The shaded region shows the upper bound set by the lowest stellar mass in our sample ($\log(M_\star/M_\odot)=10.1$), and vice versa for the highest stellar mass ($\log(M_\star/M_\odot)=11.7$), with the literature dust/CO compilation out to $z=3$ shown as a grayscale contour; note that local quiescent galaxies with $f_{H2}\sim1\%$ at $z=0$ are artificially high because the majority are upper limits.

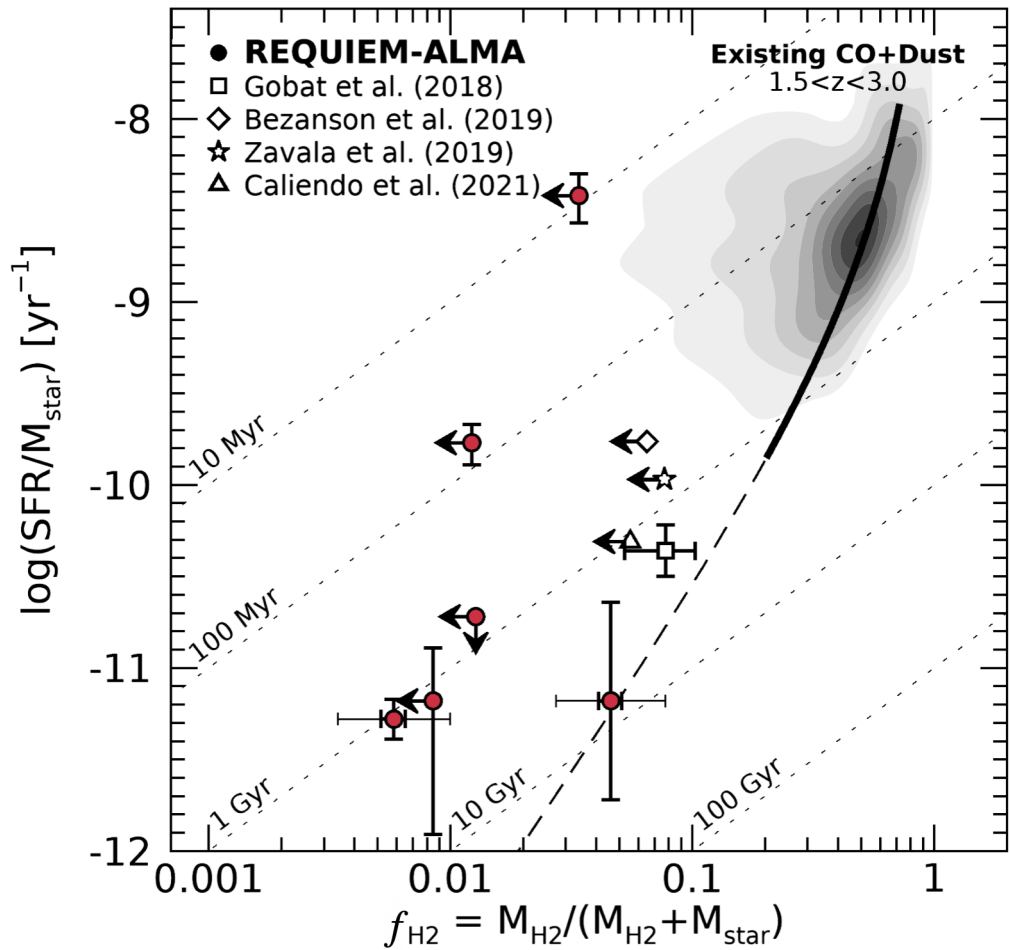

**Figure 3 | Low molecular gas masses compared to star forming galaxies.** The molecular gas fraction $f_{H2}$ is significantly lower at a given star-formation rate per stellar mass (sSFR≡SFR/M$_\star$) for distant lensed quiescent galaxies at $z>1.5$ when compared to the compilation of existing CO and dust measurements of similarly-massive star-forming galaxies (contours, see Methods). Our sample explores an order of magnitude lower sSFR and higher redshifts, finding median molecular gas fractions a factor of 10 lower than existing measurements for distant quiescent galaxies[4-6,8] (see Methods). The thick horizontal error bars for the two new detections represent formal 1σ measurement uncertainty in our 1.3mm flux density and the thin horizontal error bars represent systematic uncertainties when varying dust temperature and molecular gas to dust ratio. Vertical error bars are 1σ uncertainties. The data are largely consistent with rapid gas depletion, on average following the tracks for constant gas depletion timescales of order ~1 billion years (dotted lines).

**METHODS**

**Cosmology and Initial Mass Function Assumptions.** Throughout this paper we assume a simplified cosmology of $\Omega_M$=0.3, $\Omega_\Lambda$=0.7, and $H_0$= 70 km s$^{-1}$ Mpc$^{-1}$ when calculating physical parameters. Such values are commonly assumed to make literature comparisons easier, as the precise measured values evolve over time. We adopt the Chabrier[36] initial mass function throughout, correcting literature values where appropriate.

**Hubble and Spitzer Space Telescope Observations.** The full details of the data reduction of the REQUIEM *Hubble Space Telescope* (*HST*) and *Spitzer Space Telescope* data are found in the REQUIEM methodology paper[37]. All targets have a minimum of 5 (up to 16) *HST* and 2 *Spitzer* Infrared Array Camera filters, covering $\lambda_{rest}$~1000Å to ~1μm. In addition to ground-based spectroscopic campaigns[15-18], *HST* Wide-Field Camera 3 G141 grism spectroscopy exists for five out of the six targets, excluding MRG-M1423.

**Star-formation Rate and Stellar Mass Estimates.** Star formation rates and stellar mass estimates are taken from the literature, derived from joint analyses of photometry and ground-based spectroscopy, modeling the rest-frame ultraviolet to near-infrared spectral energy distribution[16,18]. These papers adopt the Calzetti[38] dust attenuation curve and parameterized star formation histories when fitting the stellar continuum with stellar population synthesis models[40]. Both exponentially decaying[16] and similar star formation histories that allow linear growth before the exponential decay[18] yield consistent stellar mass and star-formation rate estimates and are generally well-suited to describe quiescent galaxies[39]. The procedures to fit the data to stellar population models marginalize over the spectroscopic redshift, velocity dispersion, age, metallicity, dust attenuation, and the emission line parameters, including an analysis of systematic uncertainties introduced by the model assumptions.

The choice of dust attenuation law and SFH will impact the inferred stellar masses and SFRs in particular. Studies suggest that the dust attenuation law is not universal[41] and that the Calzetti attenuation curve may not be representative at high redshift and/or low sSFR[42]. Moreover, parametric SFHs are shown to yield systematically lower stellar masses owing to younger ages relative to non-parametric SFHs[43]. While there remains significant uncertainties in the dust geometry given our centrally-concentrated, unresolved dust continuum detections, it is valuable to test the impact of different dust attenuation assumptions and a non-parametric star formation history on the global measurement of stellar mass and sSFR through a preliminary joint analysis of the *HST* and *Spitzer* photometry and *HST* grism spectroscopy using published Bayesian methodology[37]. Namely, we independently derive the measured stellar masses and sSFRs for the two ALMA-detected galaxies, MRG-M0138 and MRG-M2129. We adopt non-parametric SFHs, using the FSPS models[44] with a 2-parameter dust model[45]. Consistent with expectations, these tests yield higher stellar masses by 0.1-0.2 dex, and thus lower implied molecular gas and dust fractions. When including the 1.3mm measurement in the fit for MRG-M0138, we find that: (1) the (non-parametric) SFH is exponentially declining, with old ages, low sSFRs, and low dust consistent with the ground-based spectroscopic results, and (2) the dust temperatures are preferentially warmer. A warm luminosity-weighted dust temperature of ~34K is required to explain the low ALMA flux density, as most of the IR energy is escaping at shorter wavelengths. On the other hand, for MRG-M2129 we find: (a) a younger post-starburst SFH with moderate dust attenuation and a steeper than Calzetti curve, and (b) the dust spectral energy distribution (SED) is consistent with a very cold (~14K), though not maximally cold, temperature. Combined, these implied dust temperatures fall at the extremes of local observations for early-type galaxies (see Molecular Gas Mass Estimates sub-section below), and make testable predictions motivating future observations. However, it is important to note that the dust is only coming from a yet unconstrained small central region, making it imperative to not over-interpret the global SED modeling.

Regardless of the specific model adopted, the changes in stellar mass and sSFR for these galaxies do not impact the conclusions of this study, despite the significant challenges of constraining the SFR in the low sSFR regime in particular. If anything, our tests imply even more extreme cold gas depletion timescales of order 100 Myr (versus ~1 Gyr for previous SED modeling assumptions). So while we conservatively adopt the published values based on higher resolution ground-based spectroscopy[16,18], derived completely independent from the dust masses, we note that our measurements may deviate even further from scaling relations under different modeling assumptions, which would only serve to strengthen our conclusions.

**Lens Model Assumptions.** The full details of the lens models for all strong lensed sources presented herein can be found in the original discovery papers[15-18]. The magnification factor was used to correct the stellar masses and star-formation rates. However, because the dust and molecular gas fractions and the specific star-formation rates, the main focus of this paper, are relative quantities, they are independent of the details of the lens models.

**Reduction of ALMA Data.** ALMA 1.3mm continuum observations were carried out in programs 2018.1.00276.S and 2019.1.0027.S. The observations were designed to reach limits on $f_{H2}$~1%; due to the range in redshift and lensing magnification within the sample, the observations reach 1σ depths of 9-56μJy. The correlator was configured for standard Band 6 continuum observations, with 7.5GHz total usable bandwidth. The data were reduced using the standard ALMA pipeline and imaged with natural weighting to maximize sensitivity. The observations were designed to avoid spatially resolving the target sources to the extent possible, and reach spatial resolutions ~1.0-1.5 arcseconds. We create lower-resolution images of each source with a *uv* taper and find no evidence for extended emission in any source (see below). Flux densities for the two detected sources were measured from the peak pixel values in the images, appropriate for unresolved (pointlike) sources. For the remaining undetected sources, we place upper limits on the 1.3mm emission using the image root-mean-square values, under the assumption that the dust emission in the remaining sources would also be as compact as that in the two detected galaxies.

We further verify that the submillimeter emission is unresolved in the two ALMA-detected objects in several different ways: comparing peak to integrated flux densities, comparing asymmetric tapered to untapered fluxes (such that the position angle of the resulting synthesized beam is aligned with the extended lensed arc), and *uv* plane and image-plane fitting, all of which agree that the two detected sources are indeed pointlike, with no evidence of extended emission.

For the two detected galaxies, we carry out a further test for millimeter emission extended on scales as large as the rest-optical light. Briefly, we create a series of mock ALMA observations using a model of the image-plane stellar light from the HST/F160W images, renormalize the image to have either a known total flux density or known peak flux density (per beam), invert it, sample the Fourier transform at the *uv* coordinates of the real data, and add noise to the visibility model based on the noise properties of the real data. The two detected sources are representative of the others: MRG-M0138 is a highly extended arc, while MRG-M2129 is only slightly extended compared to the synthesized beam. By normalizing the model image to match the peak surface brightness of the real data, we test the extent to which the existing data rule out millimeter emission with comparable extent as the stellar light. By normalizing the model image to a known total flux density, on the other hand, we test our ability to recover known input signals and whether it is possible to find pointlike millimeter emission even if the true emission has the same structure as the stellar light. In both tests, we find that if the millimeter emission had identical structure to the stellar light, the resolved arc structure of the source would have been clearly detected in both cases. We find that if the sources had the same total flux density as we measure in the real data but this emission was distributed over the full extent of the stellar light, we would still accurately recover the input flux density. Importantly, in this case the mock observed galaxies fail all of our previous tests for pointlike emission; the extended nature of the millimeter emission in the mock datasets would be easily discernible at the depth of our data. The millimeter emission in the detected sources is genuinely pointlike at our current spatial resolution, far less extended than the stellar light. Thus, we conclude that there is no significant dust emission extended on the same scales as the stellar light, in agreement with our finding that the submillimeter emission is point-like at the current resolution.

Galaxies that are not detected afford an opportunity to stack the dust continuum, reaching below the noise level for any individual map. While large variations in the strong lensing magnification coupled with small number statistics complicate matters, we generate a weighted stack as a test under the following assumptions. For the four undetected REQUIEM-ALMA galaxies, each non-detection map is divided by the magnification and the individual maps' demagnified root-mean-square defines the weight when averaging to generate a weighted stack. This methodology is similar to others in the literature for unresolved sources [46], with our sample having roughly similar beam sizes that span 1.4-1.6 x 1.1-1.2 arcseconds. The same weights are used to calculate the average stellar mass and consequently the limit in $f_{dust}= M_{dust}/M_{\star}$ for the stack. The resulting deep 3σ limit in the dust continuum from the undetected REQUIEM-ALMA sources is 4.58 μJy at an average redshift of $z$=2.59. For an average weighted stellar mass of $\log_{10}(M_{\star}/M_{\odot})$ of 10.50, this corresponds to $f_{dust}$ of $1.8 \times 10^{-4}$, largely driven by the highest magnification source, MRG-M1341, that also has the deepest ALMA 1.3mm limits but the lowest stellar mass.

**Molecular Gas Mass Estimates.** By probing the Rayleigh-Jeans tail at $\lambda_{rest}$>250μm, the dust continuum can be used as a proxy for the mass of the molecular interstellar medium, $M_{H2}$. We estimate dust mass, $M_{dust}$, from a modified blackbody fit[47], assuming a dust temperature of 25K, a dust emissivity index, β, of 1.8, and a dust mass opacity coefficient, $\kappa_{345GHz}$ of 0.0484 m²/kg[48]. By assuming a molecular gas to dust mass ratio, δ, of 100[48], we can thereby infer $M_{H2}$ from $M_{dust}$. In principle $M_{dust}$ could trace both neutral and molecular hydrogen, and quiescent galaxies at $z$~0 are known to harbor non-negligible neutral gas reservoirs[49]. Local studies show that the neutral hydrogen contribution varies widely[21,50]. While we assume that all of the hydrogen gas is in the molecular form, a significant contribution from neutral hydrogen to our dust detection would only serve to strengthen our conclusion. For comparison, we also calculate $M_{H2}$ explicitly following an empirical calibration[19], finding an offset of 0.1 dex lower in $M_{H2}$, yielding even lower inferred molecular gas fractions.

An alternative viable explanation of the null detections is that δ increases dramatically for a significant fraction of early quiescent galaxies. There exists theoretical[51] and observational[52] evidence that in certain circumstances thermal sputtering by hot electrons could in principle efficiently destroy dust in dead galaxies. CO observations are required to rule out extreme molecular gas to dust ratios that would be necessary to reconcile our observations with higher, more typical values of $f_{H2}$. While CO observations of quiescent galaxies at $z$>1.5 are scant, such ratios are difficult to justify, as they imply that CO should be detectable[7,53]. At least in the case of our two detections, such exotic ratios are already ruled out by strong CO upper limits[54].

We adopt a dust temperature of 25K, which corresponds to a luminosity-weighted temperature of roughly 30K. However, the cold interstellar medium of local quiescent galaxies is generally colder, with luminosity-weighted dust temperatures observed to be 23.9±0.8K (with a range from 16K to 32K)[55]. While adopting significantly colder dust templates would increase our estimates of molecular gas fraction, our upper limits would still leave room for tension. The thin error bars for the two detected sources in Figure 2 represent the systematic uncertainty in dust temperature from a Monte Carlo analysis adopting the observed temperature distribution of local quiescent galaxies[55]. Systematic uncertainties in Figure 3 additionally include variation in the molecular gas to dust ratio by conservatively drawing from a uniform distribution ranging from 50 to 200. Star formation in quiescent galaxies at high redshift is generally less suppressed in comparison to local dead galaxies, and as such the expected dust temperature of the cold interstellar medium remains unclear.

When including the measured ALMA 1.3mm flux density in global SED modeling that assumes energy balance, as described above, we find that MRG-M0138 may be consistent with warmer dust with luminosity-weighted temperatures of ~34K and MRG-M2129 with colder dust at ~14K. While we cannot draw conclusions on the dust temperature based on a single (unresolved) data point in the Rayleigh Jeans tail with ample uncertainties looming in the dust geometry, the latter may support conclusions based on stacked observations[9], whereas the former would represent a new extreme. It may be that once dust production from new star formation halts, dust is slowly removed by other physical processes; when the dust reservoirs are sufficiently depleted, the galaxy is optically thin and this dust may then be heated to higher temperatures. However, while the dust is still optically thick, self-shielding may effectively allow the dust to cool to very low temperatures[43]. Future observations and spatially-resolved analyses will illuminate the dust morphology and temperature.

**Literature Comparisons.** We include quiescent targets measured through dust continuum in Figure 2, both upper limits for individual galaxies[5,6] and from stacking[8,9], as well as an individual quiescent CO upper limit measurement[4] to Figure 3; all studies have a similar high redshift of $z>1.5$. For the dust continuum measurements, all data is recalibrated using the same set of assumptions applied herein, starting from the flux density and source redshift. We compare our results to a comprehensive compilation of 843 galaxies out to $z=3$ from the literature with dust or CO measurements[6], shown as contours in Figure 2. Within this sample, we highlight measurements of 188 (almost exclusively star-forming) galaxies at $1.5<z<3.0$, tracing molecular gas via dust continuum[6,19,56,57] and CO[11,12,14,58-64], comprising the contours presented in Figure 3.

**Methods References.**

36. Chabrier, G., "Galactic Stellar and Substellar Initial Mass Function", Pub. Astron. Soc. Pac., 115, 763-795 (2003).
37. Akhshik, M., et al., "REQUIEM-2D Methodology: Spatially Resolved Stellar Populations of Massive Lensed Quiescent Galaxies from Hubble Space Telescope 2D Grism Spectroscopy", Astrophys. J., 900, 184 (2020).
38. Calzetti, D., et al., "The Dust Content and Opacity of Actively Star-forming Galaxies", Astrophys. J., 533, 682-695 (2000).
39. Lee, B., et al., "The Intrinsic Characteristics of Galaxies on the SFR-M$_*$ Plane at $1.2 < z < 4$: I. The Correlation between Stellar Age, Central Density, and Position Relative to the Main Sequence", Astrophys. J., 853, 131 (2018).
40. Bruzual, G., & Charlot, S., "Stellar population synthesis at the resolution of 2003", Mon. Not. R. Astron. Soc., 344, 1000-1028 (2003).
41. Salmon, B., et al., "Breaking the Curve with CANDELS: A Bayesian Approach to Reveal the Non-Universality of the Dust-Attenuation Law at High Redshift", Astrophys. J., 827, 20 (2016).

42. Salim, S., et al., "Dust Attenuation Curves in the Local Universe: Demographics and New Laws for Star-forming Galaxies and High-redshift Analogs", Astrophys. J., 859, 11 (2018).
43. Leja, J., et al., "An Older, More Quiescent Universe from Panchromatic SED Fitting of the 3D-HST Survey", Astrophys. J., 877, 140 (2019).
44. Conroy, C., Gunn, J., & White, M., "The Propagation of Uncertainties in Stellar Population Synthesis Modeling. I. The Relevance of Uncertain Aspects of Stellar Evolution and the Initial Mass Function to the Derived Physical Properties of Galaxies", Astrophys. J., 699, 486-506 (2009).
45. Kriek, M., & Conroy, C., "The Dust Attenuation Law in Distant Galaxies: Evidence for Variation with Spectral Type", Astrophys. J. L., 775, 16 (2013).
46. Johansson, D., Sigurdarson, H. & Horellou, C., "A LABOCA survey of submillimeter galaxies behind galaxy clusters", Astron. Astrophys., 527, 117 (2011).
47. Greve, T., et al., "Submillimeter Observations of Millimeter Bright Galaxies Discovered by the South Pole Telescope", Astrophys. J., 756, 101 (2012).
48. Scoville, N., et al., "The Evolution of Interstellar Medium Mass Probed by Dust Emission: ALMA Observations at z = 0.3-2", Astrophys. J., 783, 84 (2014)
49. Zhang, C., et al., "Nearly all Massive Quiescent Disk Galaxies Have a Surprisingly Large Atomic Gas Reservoir", Astrophys. J. L., 884, 52 (2019).
50. Sage, L., et al., "The Cool ISM in Elliptical Galaxies. I. A Survey of Molecular Gas", Astrophys. J., 657, 232-240 (2007).
51. Li, Q., et al., "The dust-to-gas and dust-to-metal ratio in galaxies from z = 0 to 6", Mon. Not. R. Astron. Soc., 490, 1425-1436 (2019).
52. Smercina, A., et al., "After the Fall: The Dust and Gas in E+A Post-starburst Galaxies", Astrophys. J., 855, 51 (2018).
53. Morishita, T., et al., "Extremely Low Molecular Gas Content in the Vicinity of a Red Nugget Galaxy at z = 1.91", Astrophys. J., 908, 163 (2021).
54. Man, A., et al., "Low molecular gas fraction in massive quenched galaxies at z>=2" (in prep)
55. Smith, M., et al., "The Herschel Reference Survey: Dust in Early-type Galaxies and across the Hubble Sequence", Astrophys. J., 748, 123 (2012).
56. Saintonge, A., et al., "Validation of the Equilibrium Model for Galaxy Evolution to z~3 through Molecular Gas and Dust Observations of Lensed Star-forming Galaxies", Astrophys. J., 778, 2 (2013).
57. Franco, M., et al., "GOODS-ALMA: The slow downfall of star formation in z = 2-3 massive galaxies", Astron. Astrophys., 643, 30 (2020).
58. Tacconi, L., et al., "Submillimeter Galaxies at z~2: Evidence for Major Mergers and Constraints on Lifetimes, IMF, and CO-$H_2$ Conversion Factor", Astrophys. J., 680, 246-262 (2008).
59. Daddi, E., et al., "Very High Gas Fractions and Extended Gas Reservoirs in z = 1.5 Disk Galaxies", Astrophys. J., 713, 686-707 (2010).
60. Silverman, J., et al., "A Higher Efficiency of Converting Gas to Stars Pushes Galaxies at z~1.6 Well Above the Star-forming Main Sequence", Astrophys. J. L., 812, 23 (2015).
61. Decarli, R., et al., "The ALMA Spectroscopic Survey in the Hubble Ultra Deep Field: Molecular Gas Reservoirs in High-redshift Galaxies", Astrophys. J., 833, 70 (2016).

62. Rudnick, G., et al. "Deep CO(1-0) Observations of $z = 1.62$ Cluster Galaxies with Substantial Molecular Gas Reservoirs and Normal Star Formation Efficiencies", Astrophys. J., 849, 27 (2017).
63. Spilker, J., et al., "Low Gas Fractions Connect Compact Star-forming Galaxies to Their $z\sim2$ Quiescent Descendants", Astrophys. J., 832, 19 (2016).
64. Aravena, M., et al., "The ALMA Spectroscopic Survey in the Hubble Ultra Deep Field: The Nature of the Faintest Dusty Star-forming Galaxies", Astrophys. J., 901, 79 (2020).

**Data Availability Statement:** Data that supports the findings of this study is publicly available through the ALMA Science Archive under project codes 2018.1.00276.S and 2019.1.00227.S and the Barbara A. Mikulski Archive for Space Telescope under project code HST-GO-15663 (including additional archival data from project codes HST-GO-9722, HST-GO-9836, HST-SNAP-11103, HST-GO-11591, HST-GO-12099, HST-GO-12100, HST-SNAP-12884, HST-GO-13459, HST-SNAP-14098, HST-GO-14205, HST-GO-14496, HST-SNAP-15132, HST-GO-15466). All HST and ALMA mosaics are publicly available at DOI:10.5281/zenodo.5009315. Derived data and codes supporting the findings of this study are available from the corresponding author upon request.